\documentstyle[12pt]{article} 
\newcommand{\beqn}{\begin{equation}}
\newcommand{\eeqn}{\end{equation}}
\newcommand{\beqnr}{\begin{eqnarray}}
\newcommand{\eeqnr}{\end{eqnarray}}

\begin{document}
\setcounter{page}{0}
\thispagestyle{empty}

\begin{flushright}    
CUPP-97/2\\
hep-ph/9708444
\end{flushright}

\vskip 30pt

\begin{center}
{\LARGE\bf Next to Minimal Higgs : Mass Bounds and Search Prospects}
\vskip 5pt
{\it Anindya Datta}$^{~1, \heartsuit}$,  
{\it Amitava Raychaudhuri}$^{~2, \dagger}$

{\footnotesize $^1$Physics Department, Taki Government College, 
\\Taki, North 24 Pgs. 743429, India.\\} 
{\footnotesize $^2$Department of Physics, University of Calcutta, 
\\92 Acharya Prafulla Chandra Road, Calcutta 700009, India.}

\vskip 5pt
{\large\bf ABSTRACT} 
\end{center}
\sl
The Standard Model of electroweak interactions has one scalar doublet.
The minimal extension of this sector is effected by adding a neutral,
singlet scalar field. Depending on whether the singlet field has a
non-zero vacuum expectation value, $x$, or not, the scenario has quite
distinctive predictions.  In particular,  $x \neq 0$ produces a mixing
between the usual $SU(2)$ doublet and the singlet, giving rise to two
physical states  and a goldstone boson with non-vanishing coupling to
these.  Presence of this coupling modifies the $2 jets + {E\!\!\!/}$
signal of the Bjorken process at LEP. We update the bounds on the Higgs
mass using the LEP-1 data.  We then explore, using parton-level Monte
Carlo event generators, the production of these scalars at the LHC {\it
via} gluon-gluon fusion and subsequent detection. We compare the
signals with the expected backgrounds.

\vskip 60pt

July 1997\\
{\footnotesize\em Electronic addresses:}
$^\heartsuit$anindya@cubmb.ernet.in; $^\dagger$amitava@cubmb.ernet.in

\newpage
\section{INTRODUCTION}
\rm 
Perhaps the most challenging problem facing the particle physics
community today is the observation and understanding of the Higgs
boson, which plays the key role of breaking the $SU(2) \times U(1)$
symmetry of electroweak interactions. Efforts to find this
elementary scalar have so far proved fruitless and the absence
of the Bjorken process Higgs signal at LEP-1 sets a lower mass
limit of 60.2 GeV \cite {L3,OTHER,L2}.

Different extensions of the Higgs structure of the Standard
Model (SM) have been examined in the literature \cite{hunter}.  Of
these, the minimal is one where a neutral, $SU(2)$ singlet scalar field
is included along with the doublet.  Such singlets are common  in
superstring inspired models like $E_6$ \cite{E6}, in composite models
\cite{Pati}, in the next to minimal supersymmetric standard model
\cite{NMSSM}, {\it etc.}  The neutral singlet scalar, $N$, 
cannot directly interact with the standard
quarks and leptons but couples to the SM doublet  through quartic terms
in the potential. In the event that $N$ develops a vacuum expectation
value ({\it vev}) the spontaneous breakdown of a global $U(1)$ symmetry
is triggered off and a Goldstone mode, $J$, emerges. This boson is
the `Majoron' in models where lepton number is identified with the
$U(1)$ symmetry \cite{majoron}.  Further, the real part of the  singlet
mixes with the neutral member of the doublet.  Thus there are two
physical scalars $H_1$ and $H_2$ both of which couple to the other
members of the SM. They can be produced either at $e^+ e^-$ colliders by
the Bjorken process or at a hadron collider by gluon-gluon fusion.
Both have non-vanishing coupling  to $J$, resulting in an enrichment
of the decay modes. In particular, the decay $H_i \rightarrow J J$ is
an invisible mode which affects the usual analyses for Higgs scalar
mass limits \cite{inv1,inv2}.

In the present work, we first consider the situation where $N$ has no
vacuum expectation value. Even in this case, a quartic term in the
scalar potential yields a coupling that drives the decay $H
\rightarrow N N$. The resultant effect on the Higgs mass bound from
LEP-1 is first investigated. Next we turn to the case where $N$ has a
non-zero {\it vev}. We update the mass bounds \cite{BB} on scalars in
such a scenario using the LEP-1 results on Higgs search following the
published data from the L3 group \cite{L3}. We then move on to the
production and decay scenarios of the scalars of this model at hadron
colliders.  Their production through gluon-gluon fusion, though similar
to that in the SM, carries an additional mixing angle suppression.
Among the various decay modes, we emphasize the decay of the heavier
scalar into a pair of the lighter variety and the subsequent decay of
these into either $b \bar b$ or into $J$s thus resulting in $4b$ or
$2b + E\!\!\!/_T$ final states. The main SM backgrounds to these
processes are, respectively, the hadronic $4b$ events and $2b + gluon$
final states where the gluon jet is outside the rapidity coverage of
the detector, giving rise to missing $E_T$. We have estimated the
signal and the SM backgrounds using parton-level Monte Carlo event
generators.

The plan of this article is as follows. In section 2, we discuss the
scalar potential of the model and the resulting mass matrix. We also
relate the parameters of the potential with the masses of the physical
states and the mixing angle.  In section 3, we begin with the case
where $N$ has no {\it vev} and in the next section incorporate a
nonzero {\it vev} of $N$ and update the bounds on the scalar masses
using the LEP-1 data.  In section 5, we turn to the LHC and examine the
production rates and decay branching ratios of the scalars as well as
their signals and possible backgrounds. The conclusions are in section
6.

\section{THE SCALAR POTENTIAL AND THE RELEVANT COUPLINGS}

The most general scalar potential, invariant under $SU(2) \times U(1)$,
containing the doublet $\Phi$ and the neutral, singlet field $N$
can be written as:
\beqn
{\cal V} = \mu ^2 \Phi ^{\dagger} \Phi +  m^2 N^{*}N
  + \lambda (\Phi ^{\dagger} \Phi)^2 + \lambda {^\prime}(N^{*}N)^2
  + \xi \Phi ^{\dagger} \Phi N^{*}N
\label{scalpot}
\eeqn
The potential has an additional $U(1)$ symmetry corresponding to the
phase of $N$. All the other particles are neutral under this $U(1)$.
(In Majoron models, this symmetry is identified with lepton number.)

To start with, consider the case where the singlet field does not
acquire any {\it vev}. $SU(2) \times U(1)$ symmetry forbids any
coupling of $N$ to the standard quarks, leptons and gauge bosons.  It
only interacts with the doublet scalar through the last term in eq.
({\ref{scalpot}) which can result in the decay $H \rightarrow N N$,
where $H$ is the physical Higgs boson. The decay products for this mode
will be invisible. We discuss the details in the next section.

Next, consider the situation where the {\it vev} of $N$ is non-zero.
Let  $v/\sqrt{2}$ and $x/\sqrt{2}$ be the {\it vev}s -- both chosen
real -- for the doublet and the singlet fields respectively.  $v$ is
related to the $W$ boson mass by the relation $m_{W}$ = ${\frac {1}
{2}}g v$. $x$ breaks the $U(1)$ symmetry alluded to above and the
imaginary part of $N$ becomes a goldstone field $J$.  The ratio $tan
\beta = (x/v)$ will be a free parameter of our analysis. The  quartic
term $\xi \Phi^{\dagger} \Phi N^{*}N$ and the {\it vev} of the singlet
scalar field generates a mixing between $H_d$ and $H_s$, the real parts
of the neutral component of $\Phi$ and $N$. 

Minimisation of the scalar potential demands,
\beqn
\mu^2 = - \lambda v^2 - \frac{1}{2} \xi x^2; \;\;\; {\rm and} \;\; m^2
= - \lambda^{\prime} x^2 - \frac{1}{2} \xi v^2
\eeqn
The mass matrix of the real scalar fields is:
\beqn
M^2 =  \pmatrix{2 \lambda v^2 & \xi x v  \cr \xi x v &  
2 \lambda ^{\prime} x^2} 
\label{mmatrix}
\eeqn
The mass eigenstates $H_1$ and $H_2$ are written as:
\beqn
H_1 = H_d \cos ~\theta + H_s \sin ~\theta;\;\;\;\;{\rm and}\;\;\;
H_2 = - H_d \sin ~\theta + H_s \cos ~\theta  
\label{H1H2}
\eeqn
where
\beqn
\tan 2\theta = \frac{\xi x v}{\lambda v^2 - \lambda ^{\prime} x^2} 
\label{thet}
\eeqn
The remaining parameters of the scalar potential can be expressed in
terms of the masses $m_i$ of $H_i$, the mixing angle $\theta$ and
$\tan \beta$:
\beqn
\lambda = \frac {g^2}{8m_W^2} (m_1 ^2 ~\cos^2\theta + m_2 ^2
~\sin^2\theta)
\eeqn
\beqn
\lambda ^\prime =  \frac {g^2}{8m_W^2 \tan^2 \beta} (m_1 ^2
~\sin^2\theta + m_2 ^2 ~\cos^2\theta)
\label{lprim}
\eeqn
\beqn
\xi = \frac { g^2 (m_1 ^2 - m_2 ^2) \sin 2\theta} {8 \, m_W ^2 \, \tan
\beta}
\eeqn
A check is applied to ensure that none of these couplings exceeds
unity.

The couplings of the SM fermions with $H_i$ are the same as the SM
Higgs-fermion ones, save appropriate mixing angle factors. But in this
model $H_i$ also couples to a pair of $H_j$ ($i, j = 1,2; \,\, i \neq
j$) and to the $J$s. As these are of particular relevance, we write
them here explicitly.
\beqn
{\cal {L}}_{H_1 J J} =  -(\lambda^{'} x \sin \theta
 +  \frac{1}{2} \xi v \cos \theta)H_1 J J  
\eeqn 
\beqn
{\cal {L}}_{H_2 JJ} =  -(\lambda^{'} x \cos \theta
 -  \frac{1}{2} \xi v \sin \theta)H_2 J J  
\label{h2jj}
\eeqn
\beqnr
{\cal {L}}_{H_1 H_2 H_2} & = & -\left[ 3 \, \cos \theta \, 
\sin^{2} \theta \,v \, \lambda + 
3 \, \cos^{2} \theta \, \sin \theta \, x \, \lambda^{\prime} + 
\frac{1}{2} \cos^{3} \theta \, v \, \xi  \right. \nonumber \\ 
& & \left. - \, \cos \theta \, \sin^{2} \theta \, v \, \xi    
-  \, \cos^{2} \theta \, \sin \theta \, x \, \xi  + 
\frac{1}{2} \sin^{3} \theta \, x \, \xi \right] H_1 H_2 H_2
\eeqnr
\beqnr
{\cal {L}}_{H_2 H_1 H_1} & = & -\left[ - 3 \, \sin \theta \, 
\cos^{2} \theta \,v \, \lambda + 
3 \, \sin^{2} \theta \, \cos \theta \, x \, \lambda^{\prime} - 
\frac{1}{2} \sin^{3} \theta \, v \, \xi  \right. \nonumber \\ 
& & \left. + \, \sin \theta \, \cos^{2} \theta \, v \, \xi    
-  \, \sin^{2} \theta \, \cos \theta \, x \, \xi  + 
\frac{1}{2} \cos^{3} \theta \, x \, \xi \right] H_2 H_1 H_1
\eeqnr

\section{MASS BOUND FROM LEP-1: $x = 0$ CASE}

As noted in the previous section, if the {\it vev} of $N$ is zero then
its only interaction is with the $\Phi$ through a quartic term in the
potential -- see eq. (\ref {scalpot}). If $H$ is the physical Higgs
field after spontaneous breakdown of the $SU(2) \times U(1)$ symmetry
then the above coupling drives the decay of $H$ to a pair of $N$s, if
kinematically allowed. (It needs to be borne in mind that $N$ is a {\it
complex} neutral field.) The singlet scalar cannot couple to the quarks,
leptons, and gauge bosons. Thus Higgs decay in this channel leaves no
signature in the detector and can only be surmised from the momentum
imbalance of the final state particles. At an $e^{+} e^{-}$ collider,
when the doublet Higgs is produced along with an off-shell $Z$-boson by
the Bjorken process, only the decay products of the latter will be seen
in the detector. In particular, a hadronic $Z$-decay produces a final
state with two jets and missing energy from the Higgs decay which
mimics the SM process where the Z decays invisibly ({\it i.e.} to
neutrinos) while the Higgs decays hadronically.  With this additional
mode in the picture, the bound on the Higgs mass from the LEP-1 data is
modified.  The L3 collaboration \cite{L3} has searched for the di-jet +
$E\!\!\!/$ signature in a data sample of 3.05 million hadronic $Z$
decays and from its absence has imposed a 95\% C.L.  lower limit of
60.2 GeV \cite{L3} on the SM Higgs mass.  We can translate this bound
for the case under discussion  by incorporating in the analysis the
additional $H \rightarrow NN$ decay mode.  Instead of using a detailed
Monte Carlo event generator and applying the appropriate cuts used by
the L3 group, we demand that:
\beqnr
[\Gamma (Z \rightarrow \nu \bar {\nu} H)B.R.(H \rightarrow b \bar {b},
c \bar {c})
\epsilon_{vis} + \Gamma (Z \rightarrow (q \bar {q})
H)B.R.(H \rightarrow N N) \epsilon_{invis}] \nonumber \\ 
< 3 \Gamma (Z \rightarrow hadrons)/3.05 \times 10^{6}
\eeqnr
In the above, the $\epsilon$s are the efficiencies of the topological
cuts used by  L3 to differentiate between the signal and the
background. Since no excess over background has been observed, the 95\%
C.L. limit corresponds to a signal of 3 events.  The Z-bosons produced
at LEP-1 are predominantly transverse and in the derivation of $\Gamma
(Z\rightarrow \nu \bar {\nu} H)$ or $\Gamma (Z
\rightarrow b \bar {b} H)$ we sum over only these polarisation states
of the parent $Z$.

We present our results in fig. 1 for three representative values of the
quartic coupling constant $\xi$, namely, $\xi =$ 1 (solid curve), .0001
(large-dashed curve), and 0 (small-dashed curve), and $m_N$ = 20 (1a)
and 0 (1b) GeV. Of course, in the above, $\xi $ = 0 is just the SM
limit. The lower bound on $m_H$ from eq. (13) corresponds to the point
of intersection of the curves with the abscissa. From the figure it is
seen that this limit is almost insensitive to $m_N$ and increases with
the increase of $\xi$, with $m_H > 66.0$ GeV for $\xi = 1$, reducing to
the SM value in the $\xi = 0$ limit. The curves for all values of $\xi
$ in the range 1 to 0.01 are practically coincident since for these
cases the branching ratio for $H \rightarrow NN$ is essentially unity.
Only when $\xi $ gets smaller, do the other modes become comparable.
(The L3 group also briefly discusses the search for an invisibly
decaying Higgs particle without going into details of any particular
model \cite{L3}, and a 95\% C.L.  lower limit on the mass of such a
scalar is quoted as 66.7 GeV -- similar to the one in the $\xi =1$
case.)

At this point it will be meaningful to compare the branching ratio in
this model of some decay channel (any channel important in the context
of SM Higgs search) with that in the Standard Model.  As an example,
the ratio of the corresponding BR's for the case of $\gamma \gamma$
decay is presented in fig. 2 as a function of $m_H$ for different
values of the parameter $\xi$.  Though the partial width for $H
\rightarrow \gamma \gamma $ is unchanged in this case, the total width
has an additional contribution from $H \rightarrow NN$. The exhibited
ratio is thus just $\Gamma(H)_{SM}/\Gamma(H)$.

\section{MASS BOUND FROM LEP-1: $x \neq 0$ CASE}

As already noted in section 2, when the singlet scalar has a nonzero
{\it vev} there are two physical massive scalar fields $H_1$ and $H_2$
and a Goldstone boson, $J$. If one or both of the $H_i$ are lighter
than the $Z$ then they will be produced {\it via} the Bjorken process
at LEP-1 so that the data can be used to set a bound on their masses.
For small values of $\tan \beta$, both have considerable decay
branching ratios to a pair of $J$s, an invisible decay mode. Thus,
once produced by the Bjorken process, they can give rise to much the
same signal as discussed in the previous section, decaying into a pair
of $J$s, and with the $Z^{0*}$ decaying to a $q \bar q$ pair,
mimicing the di-jet + $E\!\!\!/$ signal of a Higgs scalar. As in the
previous section, to derive the lower bounds on the scalar masses we
require that the constraint of eq. (13) be satisfied. Of course, in
this case the left side of the equation depends on $m_{1}, m_{2},
\theta $ and $\tan \beta$. As already noted, there are two
possibilities: (a) when both $m_1, \,\, m_2 < m_Z$ and (b) when only
one of the scalars is lighter than the $Z$ and contributes.

We let the mixing angle $\theta $ vary from 0 to $\pi /2$. It is
readily seen that the range $\pi /2 \leq \theta \leq \pi $ can be
mapped to the previous one by the  redefinition of $H_1 \rightarrow
H_2$ and $H_2 \rightarrow - H_1 $ -- see eq. (\ref {H1H2}).

The situation dealt with in the previous section corresponds to $\tan
\beta = 0$. Here we let it vary in the range $1 \leq \tan \beta \leq
10$ which corresponds to a maximum value of $x \simeq 2$ TeV.

We present in fig. 3 our results for the case where both scalars are
lighter than $m_Z$ for two representative values of $\tan \beta$,
namely 1 (solid line), and 10 (dashed line). Three choices of the
mixing angle -- {\it viz.} $\theta $ = $5^{o}$, $45^{o}$, and $85^{o}$
-- are considered. The allowed masses lie in the region to the right of
the curves. From the figure it is seen that the lower mass limit can be
substantially low for small or large values of the mixing angle
$\theta$.  On the contrary, for the case $\theta $ = $45^{o}$ both
scalars are heavier than about 55 GeV.  We have not exhibited the
region below a mass of 20 GeV. In fact, a small window survives in the
mass range of 5 -- 10 GeV. This is an artefact of the low efficiency of
detection for this mass range.

If one of the scalars has a mass more than $m_Z$ then only the other
one can contribute to the Bjorken process. The results in this case,
for the same representative values of $\tan \beta$, namely 1 (solid
line), and 10 (dashed line), are shown in fig. 4.  The allowed masses
and mixing angles are to the right of the curves.  When $H_2$ is the
lighter scalar, the mass bound must reduce to the SM value in the
$\theta = 90^o$ limit, independent of $\tan \beta $, in consonance with
eq. (\ref{H1H2}).  (If $H_1$ is the lighter one, then the corresponding
limit is $\theta = 0^o$.) Our findings agree with that of the L3 group
\cite{L3} who obtain a bound of 58 GeV for the SM Higgs using only the
invisible decay modes. They also quote a stronger bound of 60.2 GeV,
but this utilises other $Z$-decays ({\it e.g.} to charged leptons) as
well, and consequently is not of relevance to this discussion.  Notice
that depending on the mixing angle, in this model a light scalar is
still allowed by the data. We have checked that these limits are
insensitive to the mass of the heavier scalar.

With a larger integrated luminosity available, for the same choice of
parameters the bounds are now strengthened compared to earlier work
\cite{BB}. In this analysis, for the $x \neq 0$ case, the mixing angle
$\theta$ can be zero only if $\xi = 0$ -- see eq. (\ref {thet}).  In
this limit the state $H_1$ is purely doublet and also has no coupling
to $JJ$, independent of $\tan \beta$. This is borne out in Fig. 4 by
the coming together of the curves for different $\tan
\beta $ in this limit. Near $\theta = 0$, the state $H_2$ is almost
purely singlet -- at $\theta =0$ this scalar cannot be produced in the
Bjorken process at all -- and its decay to the invisible $JJ$ final
state is controlled by the product $\lambda ' x$ -- see eq. (\ref
{h2jj}) -- which is inversely proportional to $x$ (note eq.
(\ref{lprim})). Consequently, for small $\theta $ the bound on $H_2$ in
fig. 4 is weakened for larger $\tan \beta$. The $\theta = \pi/2$ case
is very similar except that $H_1$ and $H_2$ are interchanged.

\section{PRODUCTION AND DETECTION AT THE LHC} 

Now we turn to the production and detection possibilities of the
scalars of this model at hadron colliders, specifically the LHC. 

For the $x = 0$ case, there is no mixing between the doublet and the
singlet scalars. The production from gluon fusion remaining unchanged,
the only difference from the SM will be in the branching ratios of the
Higgs scalar since new decay modes become available. All SM decay
channels suffer the same suppression in the BR due to the additional
invisible modes discussed in section 3.  This suppression ratio has
been plotted in fig. 2. Thus, if an SM-like Higgs is observed at the
LHC with the ratio of BRs consistent with the Standard Model but the
apparent production rate smaller than expected, then that could be
indicative of this scenario.

The $x \neq 0$ case has more novelty. Like the SM Higgs, $H_1$ and
$H_2$ can be produced at the LHC by gluon-gluon fusion \cite{BP}.  The
production rate is suppressed with respect to the SM case by
appropriate mixing angle factors.  Our stress will be on the production
of the heavier one, which we term $H$, and its subsequent decay to a
pair of the lighter -- $h$ -- ones \cite{roychou}. This decay mode, if kinematically
allowed, can give rise to new signals. We look for both the $h$s
decaying to $b \bar b$ giving rise to 4$b$ final states or one of the
$h$s decaying to $b \bar b$ and the other to a pair of $J$s, yielding
missing $E_T$ plus 2$b$ final states.  

Unfortunately, both channels, though somewhat novel, suffer from large
QCD backgrounds.  The potential source of background to the first
channel is (a) QCD 4$b$ production and (b) pair production of $Z^0$s
both of which decay in the $b \bar b$ mode. The latter is removed by
the invariant mass cut imposed on the $b$-pairs (see below). A further
contribution comes from $b\bar{b}gg$ production where the gluon jets
are misidentified as $b$-jets. The main background for the other
channel comes from the QCD 3 jet ($b \bar b g$) process,
where the gluon jet is outside the rapidity coverage of the detector
and gives rise to missing $E_T$. We have analysed these backgrounds
with parton-level Monte Carlo generators.

Before delving into the details of the analysis of the signals
(and the backgrounds) let us pay some attention to the relevant
branching ratios of interest, namely BR $(H \rightarrow b \bar b)$,
BR $(H \rightarrow hh)$, and,  finally, BR $(H \rightarrow J J)$.
Some representative results are shown in figs. 5 and 6 where $m_h$
is fixed at 70 GeV.  The convention is as follows:
\newline
solid line : $H \rightarrow b ~\bar b$  \\
large-dashed line : $H \rightarrow h h$\\
small-dashed line :$~H \rightarrow J J$ \\
For the purpose of illustration, results are presented for $\tan \beta
=1$ and $\tan \beta =10$ and mixing angles $\theta $ =  5$^o$ 
and 45$^o$. The results for $\theta = 85^o$ are almost
identical to the $\theta = 5^o$ case under the interchange $H_1
\leftrightarrow H_2$. From the plots it is seen that for the two
exhibited mixing angles the situations are vastly different and we now
discuss them in turn.

First consider the case of $\theta = 5^o$ (fig. 5).  For $H \equiv H_1$
the dominant decay mode is to a pair of $b$s.  The 3 body decay $H
\rightarrow W W^{*} \rightarrow W f
\bar {f'}$ which opens beyond the $W f \bar{f'}$ threshold is also
comparable but is not relevant for this work and has not been
displayed.  Beyond the $W^{+}W^{-}$ threshold the other branching
ratios are all very small.  On the other hand, for $H \equiv H_2$, it
is the invisible decay which is the dominant one.

For $\theta = 45^o$ (fig. 6) and $\tan \beta $ = 10, the $H \rightarrow
hh$ mode is dominant once it is kinematically allowed and the two cases
$H \equiv H_1$ and $H \equiv H_2$ are almost identical. The $\tan \beta
= 1$ and $H \equiv H_2$ case is unusual in that the $Hhh$ coupling
vanishes identically. This is not so for $H \equiv H_1$.

We have not shown the branching ratios of $h \rightarrow b ~\bar b$ and
$h \rightarrow J J$. We have checked that these BRs are independent of
$m_H$. This allows us to use the following approximate prescription:  
\[
{\rm BR} (h \rightarrow JJ)_{(\theta, m_h = 70 {\rm GeV})} = {\rm BR}
(H \rightarrow JJ)_{({\pi \over 2} - \theta, m_H = 70 {\rm GeV})}
\]
and an analogous one for the $b\bar {b}$ mode. The branching ratios for
$h$ can then be read off from figs. 5 and 6 from the extremal points
$m_H = m_h = 70$ GeV.

We now turn to the detection possibilities.  First, we look at the
$4b$ final state which has also been examined in the context of
supersymmetric Higgs boson search \cite{SUSYH}. This mode suffers from
a huge QCD background. We have analysed both the signal and the
background at the parton level using a Monte Carlo event generator. The
MRSA parton distribution functions \cite{MRS} have been used.  Signal
events consist of 4 hard $b$-jets. We have incorporated a QCD
enhancement factor ($\simeq$ 1.5) \cite{SPIRA} for the calculation of
the signal.  Conservatively, we assume the $b$-tagging efficiency is on
the average equal to 30\% \cite{LHCB}.  The main background is due to
the QCD 4$b$ process. An additional contribution comes from QCD
$b\bar{b}gg$ production, where the gluon jets are mis-tagged as
$b$-jets. We did not analyse this latter process separately, but assume
that it contributes as much as the  4$b$ background itself -- a very
safe estimate. The final state particles being almost massless with
respect to the LHC center of mass energy, the jets are boosted in the
forward direction both for the signal and the background. The following
cuts are imposed in line with the LHC detectors: \\

$p_T$ of each jet is greater than 15 GeV. \\

$|\eta|$ of each jet is less than 2.5\\ 

The kinematic distribution of the signal and the background events are
similar and the cuts are not of much help in boosting their ratio.
The main difference between them is that for the former the jets are
due to the two body decay of the scalars of lighter mass, which, in
turn, are pair produced when the heavier scalar decays. This is
incorporated in the analysis by demanding that in addition to the
constancy of the $4b$ invariant mass, it should be possible to combine
the b-jets into two pairs of the same invariant mass.

In fig. 7 the variation of the Significance ($= {S \over \sqrt B})$
with the 4-jet invariant mass -- $m_H$ -- is exhibited for different
values of the di-jet invariant mass -- $m_h$.  We present the results
for $m_h$ = 70 (7a), 100 (7b) and 130 GeV (7c).  The proposed LHC
integrated luminosity of $10^5 pb^{-1}$ for a one year run -- the high
luminosity option -- has been used. Results for  $\tan \beta = 10$ and
1 are shown and $\theta $ has been chosen to be 45$^{o}$.   

For $\theta = 5^o$ the Significance is much smaller. This can be traced
to the following: If $H \equiv H_1$ -- predominantly the doublet scalar
--  the small mixing angle suppresses the decay $h \rightarrow
b\bar{b}$ while for $H \equiv H_2$ -- essentially a singlet scalar --
the production of $H$ is very small for the same reason. In addition,
there is a further reduction due to the smallness of the $H \rightarrow
hh$ branching ratio in this case.

The notation in fig. 7 is as follows: $\tan \beta =$ 10 for the solid
and large dashed lines which correspond to $H \equiv H_1$ and $H \equiv
H_2$ respectively. The small-dashed line is for $H \equiv H_1$ and
$\tan \beta $ = 1. (The case  $H \equiv H_2$ and $\tan \beta $ = 1 does
not provide a signal, see below.) From fig. 7a  it is seen that so long
as the 4-jet invariant mass is less than twice the $W$ mass the
magnitude of the Significance is experimentally viable.  But once the
$WW$ threshold is crossed, the Significance comes down even below
unity.  This reflects the fact, mentioned earlier, that if the $WW$
decay  is kinematically allowed, it dominates over the other modes.
The number of $4b$ events from the background is also very high in this
region. As the 4-jet invariant mass is increased, the number of
background events drops sharply.  So for higher values of the 4-jet
invariant mass -- see also figs. 7b and c -- the Significance again
climbs up. At first sight, it may seem odd that the Significance
increases with $m_H$. This is actually due to the rapid fall in the
background, which more than compensates the gentle decrease of the
signal.  Notice that we do not present any results for the $\tan
\beta $ =1 and $H \equiv H_2$ case. This is because, as noted earlier,
for this case the $H \rightarrow hh$ decay is forbidden.

The Significance is seen to be higher for the $\tan \beta $ = 10 case.
The signal cross-section can be further increased by choosing a larger
$\tan \beta $. But $\tan \beta$ beyond 10, pushes the singlet {\it vev},
$x$, beyond the TeV scale, which is not very appealing.

Next, consider the case when one of the $h$ decays invisibly through
the $J J$ mode leaving a 2 $b$-jets $ + E\!\!\!/_{T}$ signal. As
pointed out earlier, the main source of background for this channel is
the QCD $b\bar{b}g$ events when the gluon jet is outside the rapidity
coverage of the detector.  Another source of background is $b \bar b
\nu \bar \nu$ production where the $\nu$s originate from the decay of
a $Z^0$.  But this latter rate is suppressed with respect to the
previous one and is ignored. As can be seen from the branching ratio
plots in figs.  5 and 6, for $\tan \beta$ = 1 the branching ratio of
the invisible decay of the lighter scalar becomes dominant over the $b
\bar b$ decay in most cases. Thus the  number of signal events for this
final state for $\tan \beta$ = 1 can be quite substantial.  But the
presence of a large amount of missing energy in the signal events
undermines any mass reconstruction strategy similar to the one used for
the $4b$ final state. In the absence of such a procedure the background
overwhelms the signal and no further results are presented for this
decay channel.

We have not discussed the direct production and decay of the lighter
scalar $h$ so far. The production rate, {\it via} gluon fusion, will be
suppressed compared to the SM rate by mixing angle factors. For the
decays, in addition to the analogous mixing angle factor suppression,
there will be a new invisible decay mode, namely $h \rightarrow JJ$.
The resultant effect is summarised, for the case $h \equiv H_1$, in
Table 1 where we list the ratio $r = \sigma(pp \rightarrow h + X
\rightarrow b \bar b + X)$ / $\sigma(pp \rightarrow h + X \rightarrow b
\bar b + X)_{SM}$ for different $m_h$, $\theta $ and $\tan \beta$.

\vskip 0.5in
\section{CONCLUSION} 

We examined the simplest extension of the SM, with an additional $SU(2)
\times U(1)$ singlet scalar field. We separately considered the cases (a)
the singlet has no {\it vev} and (b) the singlet has a non-vanishing
{\it vev}.  In the former case, there is no mixing of the doublet
scalar field with the singlet field. The only difference from the SM
arises from new decay modes of the doublet scalars to singlets via a
quartic coupling term.  In the latter case,  there are two massive
neutral scalar fields -- linear combinations of the doublet and the
singlet states -- with non-vanishing coupling to the goldstone boson,
$J$. We have updated the bounds on the masses of the scalars in both
scenarios from the LEP-1 data. 

Here, we have not considered the expectations from this model for
LEP-2. In a recent paper \cite{deC} de Campos {\it et al.} consider the
mass bounds that may be obtained from LEP-2 for an invisibly decaying
scalar.  Though their model is somewhat different (two doublets and one
singlet) it reduces to the one of this paper in a suitable limit. We
have checked, using a parton level Monte Carlo and imposing the
appropriate cuts, that our results agree with theirs in this limit.

We have investigated the detection possibilities of the scalars at the
LHC, especially for the $x \neq 0$ case, using parton-level Monte Carlo
event generators.  In our analysis we consider the heavier scalar
decaying to a pair of the lighter ones, both of which subsequently
decay either to $b \bar b$ yielding  a $4b$ final state or one of the
lighter scalars decays invisibly ({\it i.e.} to $JJ$), giving rise to a
$2b + E\!\!\!/_{T}$ signal. Both the channels suffer from large QCD
backgrounds which have also been examined.  Using a mass reconstruction
strategy for the $4b$ final state we observe that for some regions of
the parameter space these scalars are likely to be seen at the LHC.

\newpage
\parindent 0pt

{\Large {\bf  Acknowledgements}}

\vskip 20pt

AR acknowledges partial financial support from the Council of
Scientific and Industrial Research and the Department of Science and
Technology, Government of India. Both authors are grateful to Biswarup
Mukhopadhyaya for participation in the initial stages
of the work. They acknowledge fruitful discussions with Gautam
Bhattacharyya.

\newpage

\begin{center}
{\bf FIGURE CAPTIONS}
\end{center}
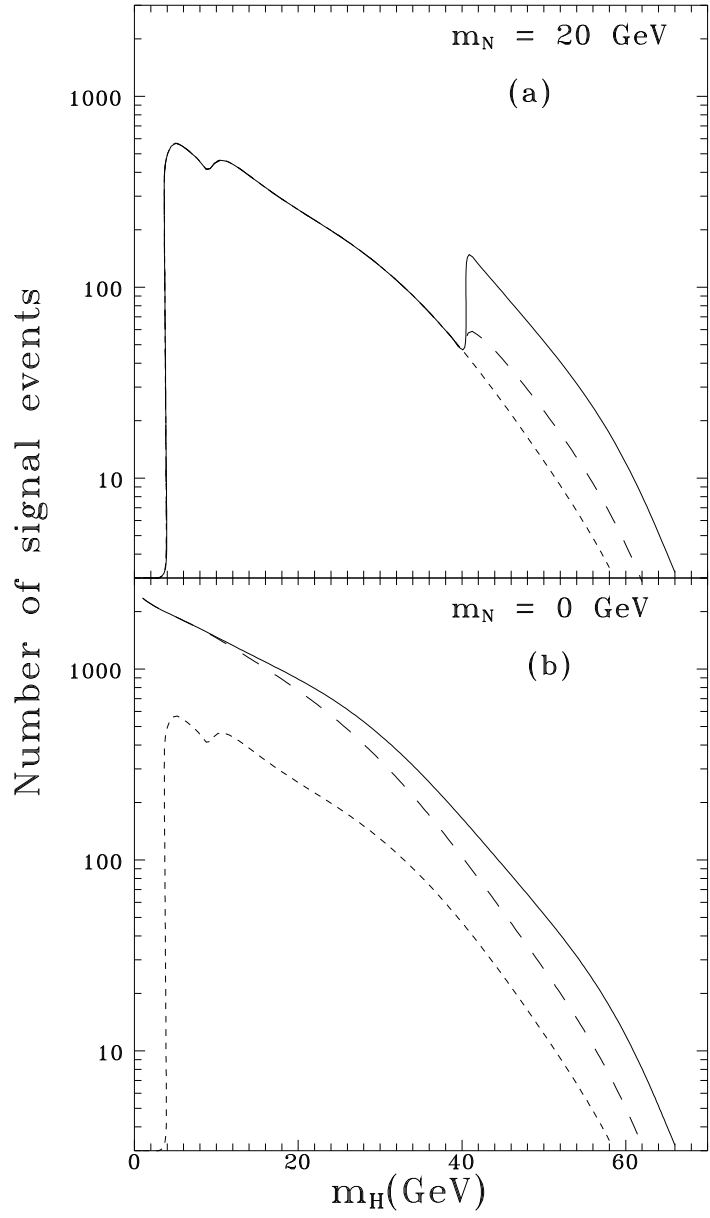
Fig.1. The number of events in the di-jet + $E\!\!\!/$ channel of
Z-decay as a function of $m_H$. The limit on $m_H$ from non-observation
of a signal corresponds to the intercept on the abscissa. The solid,
large-dashed and small-dashed curves correspond to $\xi $ = 1, 0.0001
and 0 (SM) respectively. Two choices of $m_N$ -- 20 GeV (1a) and 0 GeV
(1b) -- are shown.\\

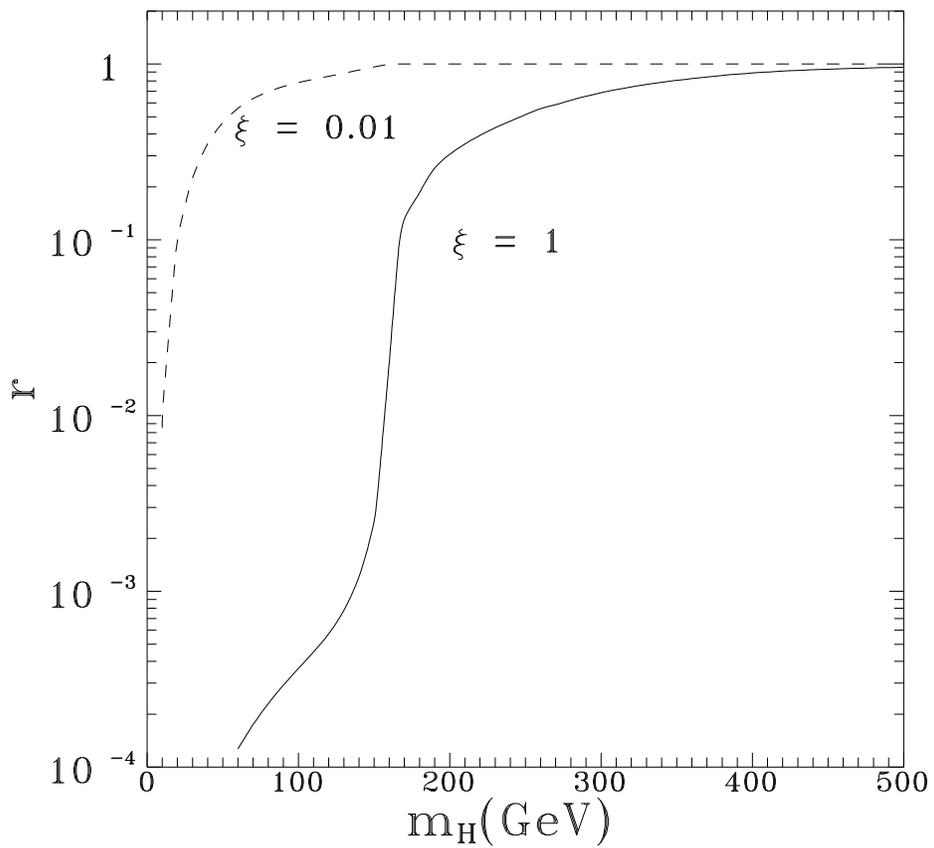
Fig. 2. The ratio $r = $ BR$(H \rightarrow \gamma \gamma)$/BR$(H
\rightarrow \gamma \gamma)_{SM}$ as a function of $m_H$ for different
values of $\xi$.\\

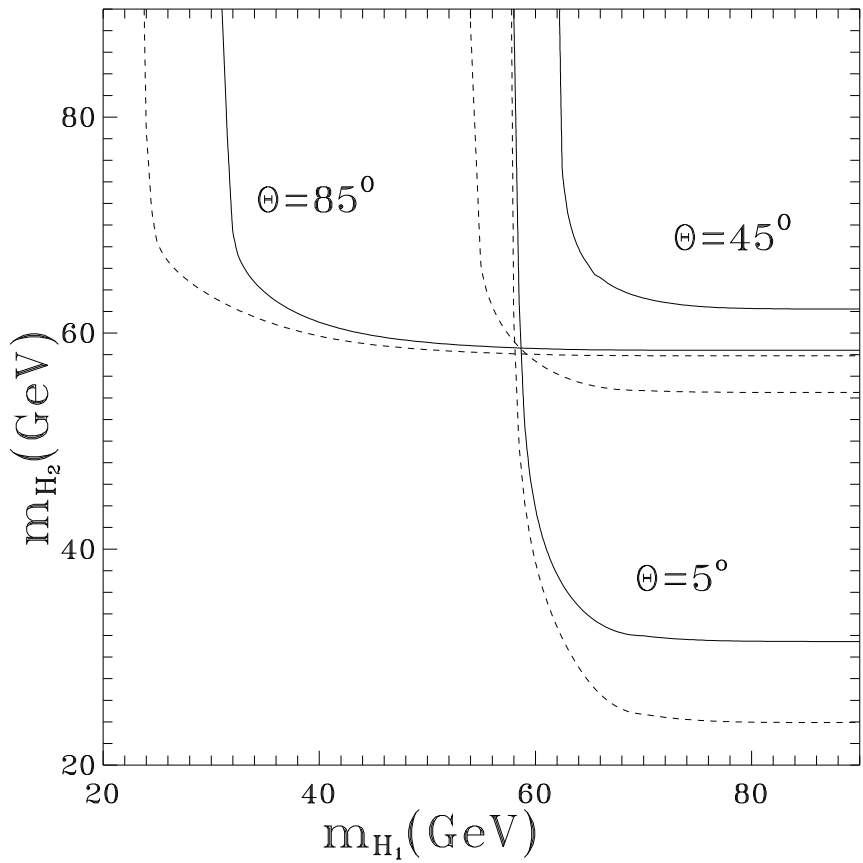
Fig. 3. The allowed mass ranges of the two scalars (to the right of the
curves) for the case $x \neq $ 0 when both are lighter than the
$Z$-boson and can be produced {\it via} the  Bjorken process. Results
are shown for two values of $\tan \beta $: 1 (solid line), and 10
(dashed line). Three choices of $\theta $ -- $5^{o}$, $45^{o}$, and
$85^{o}$ -- have been considered.\\

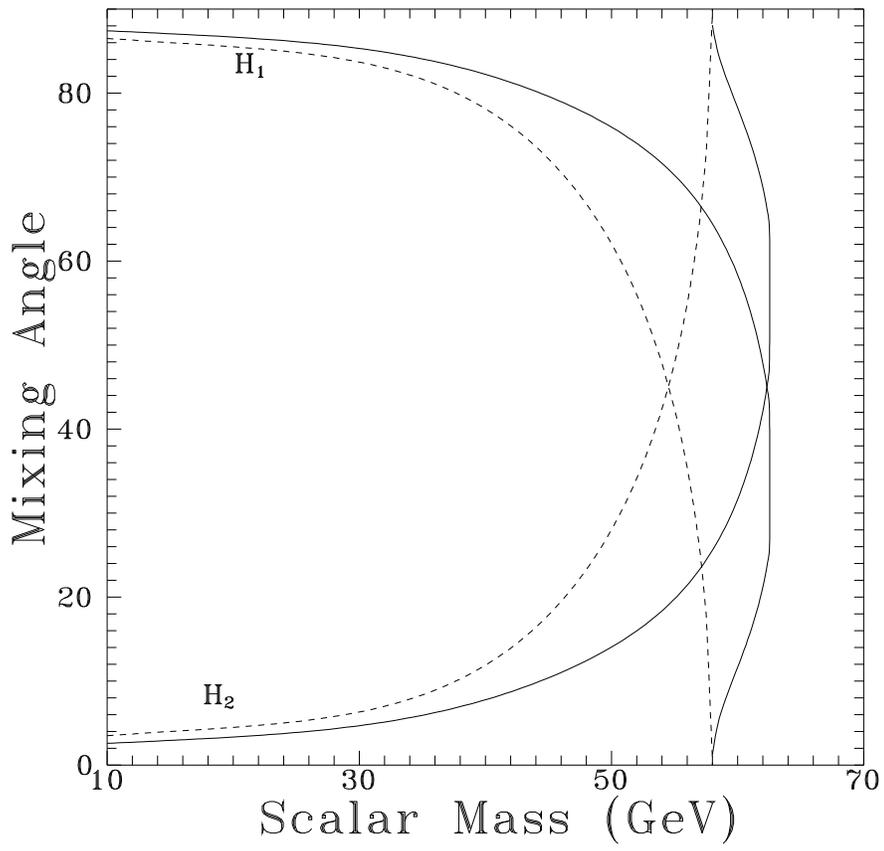
Fig. 4. The allowed mass ranges of the lighter scalar as a function of
the mixing angle for the case $x \neq $ 0 when the other scalar is
heavier than the $Z$-boson.  Results are shown for two values of
$\tan \beta $: 1 (solid line), and 10 (dashed line). The allowed
regions are to the right of the curves.\\

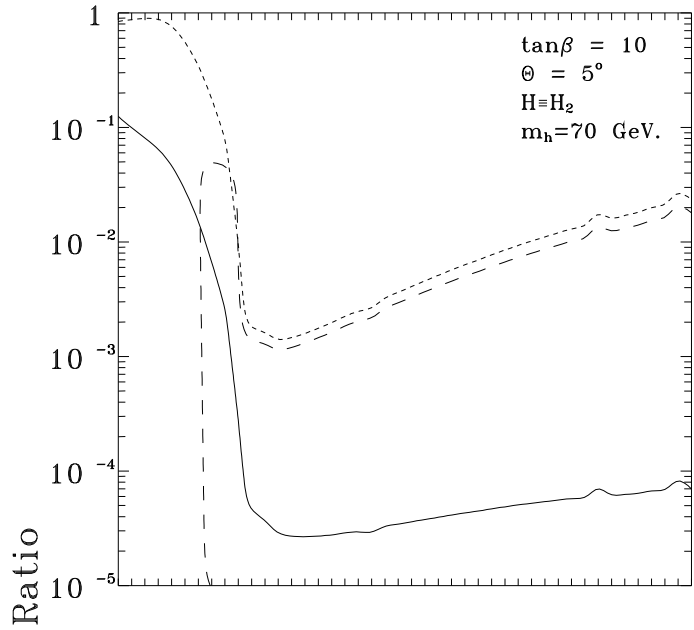
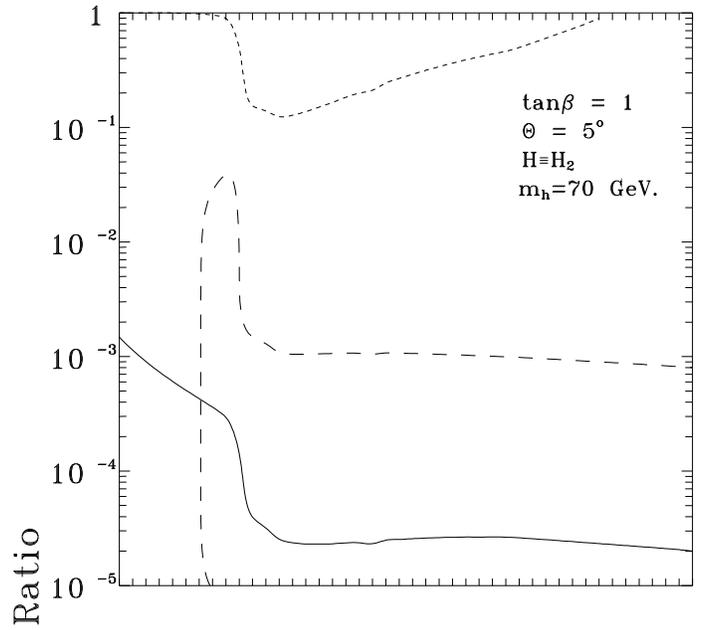
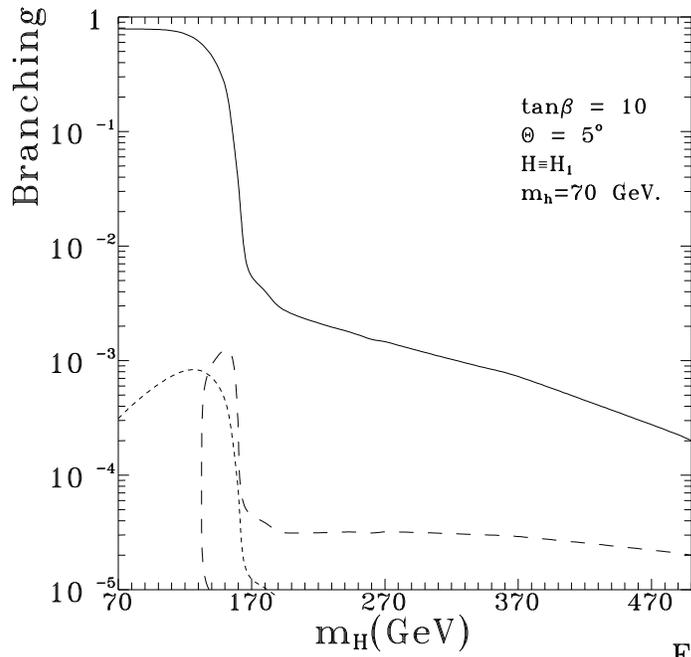
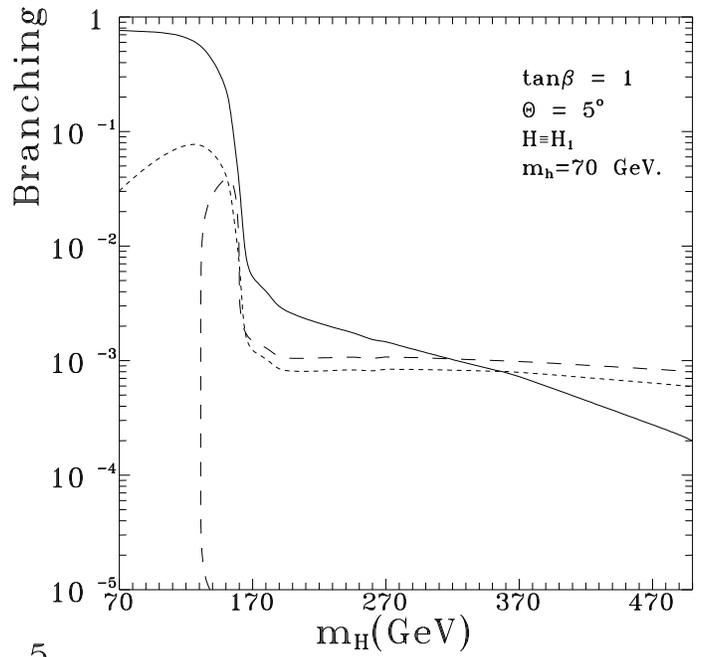
Fig. 5. Branching ratio of the heavier scalar, $H$, to decay to $b \bar
b$ (solid line), $JJ$ (small-dashed line), and $hh$ (large-dashed
line).  The two possibilities $H \equiv H_1; \,\, h
\equiv H_2,  $ and $H \equiv H_2; \,\, h \equiv H_1,  $ have been
considered. The results for two choices of $\tan \beta =$ 1 and 10 have
been shown and the mixing angle $\theta $ is chosen to be 5$^o$.\\

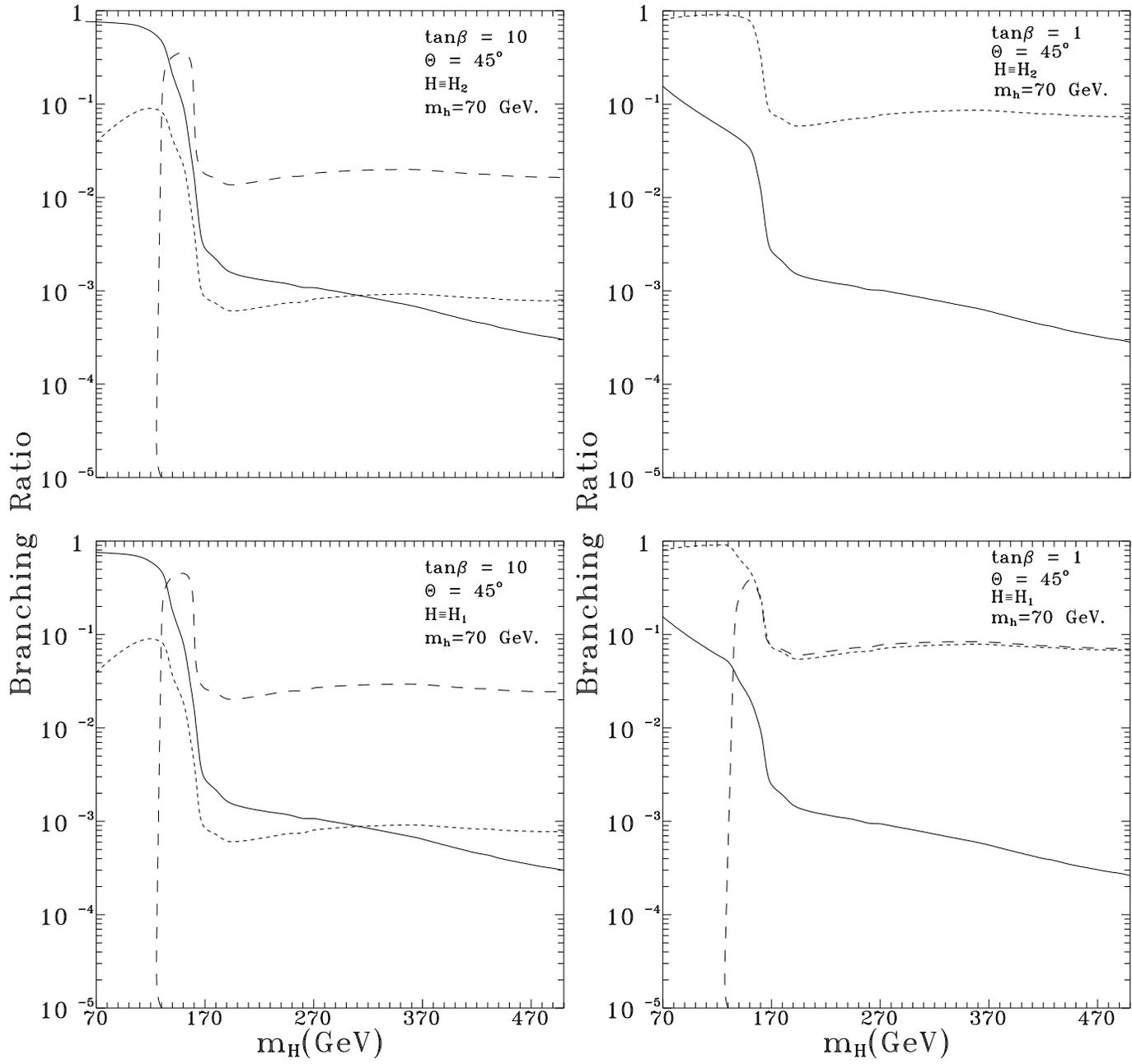
Fig. 6. Same as in fig. 5 except that the mixing angle $\theta $ is
45$^o$.\\

Fig. 7. Significance $(S/\sqrt{B})$ as a function of $m_H$ for three
different choices of $m_h$ -- 70 GeV (a), 100 GeV (b), and 130 GeV (c).
$\tan \beta =$ 10 for the solid  and large dashed lines which
correspond to $H \equiv H_1$ and $H \equiv H_2$ respectively. The
small-dashed line is for $H \equiv H_1$ and $\tan \beta $ = 1.

\vskip 40pt

\newpage
\begin{center}
{\bf TABLE CAPTION}
\end{center}
Table 1: The ratio $r = \sigma(pp \rightarrow h + X \rightarrow b \bar
b + X)$ / $\sigma(pp \rightarrow h + X \rightarrow b \bar b + X)_{SM}$,
for the case $h \equiv H_1$ for different values of $m_h$, $\tan \beta
$ and $\theta$.

\begin{center}
{\bf\large Table 1}
\end{center}

\begin{center}
\begin{tabular}{|c|c|c|c|c|}
\hline
\multicolumn{5}{|c|}{$r$} \\
\hline
{} & {} & \multicolumn{3}{c|}{$m_h$} \\ \cline{3-5}
$\theta$ &{$\tan \beta$ } &  70 GeV &
100 GeV & 130 GeV  \\ \hline
$ 5^o$ &{ 1}
& { .9624 } & {.9344} & {.9151}  \\ \cline{2-5}

{}&{10}
& {.9920}& {.9917 } & {.9900 }  \\ \hline 

$45^o$ &{ 1}
& {.0986} & {.0549} & {.0415}  \\ \cline{2-5}

{}&{ 10}
& {.4804} & {.4609} & {.3830}  \\ \hline 

$85^o$ &{1}
& $1.427 \times 10^{-5}$ & $7.166\times 10^{-6}$ & $5.27\times 10^{-6}$
\\ \cline{2-5}

{}&{10}
&{.0012} & {.0006 } & {.0002}  \\ \hline 
\end{tabular}
\end{center}

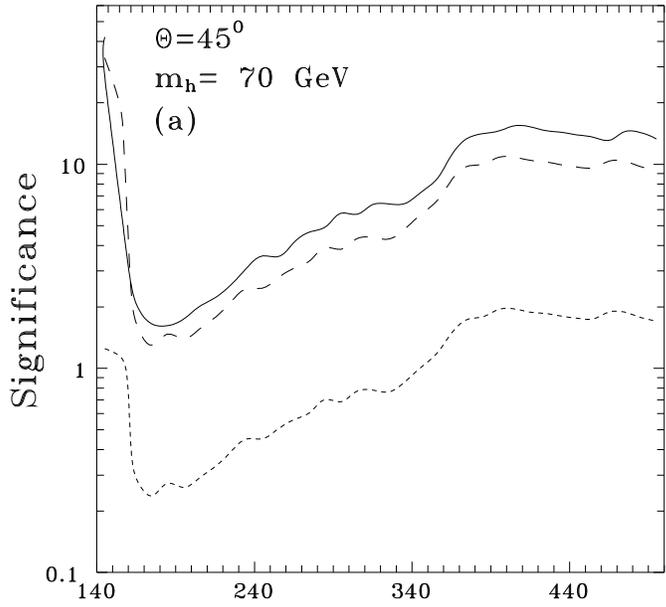
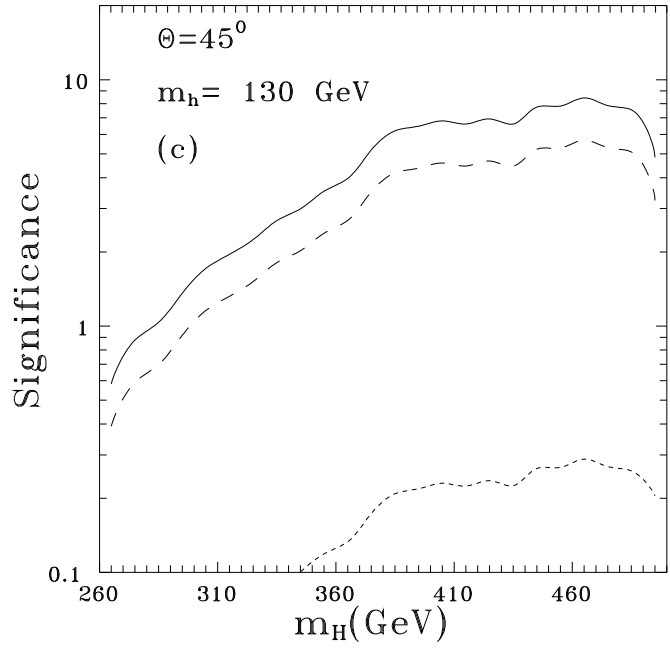
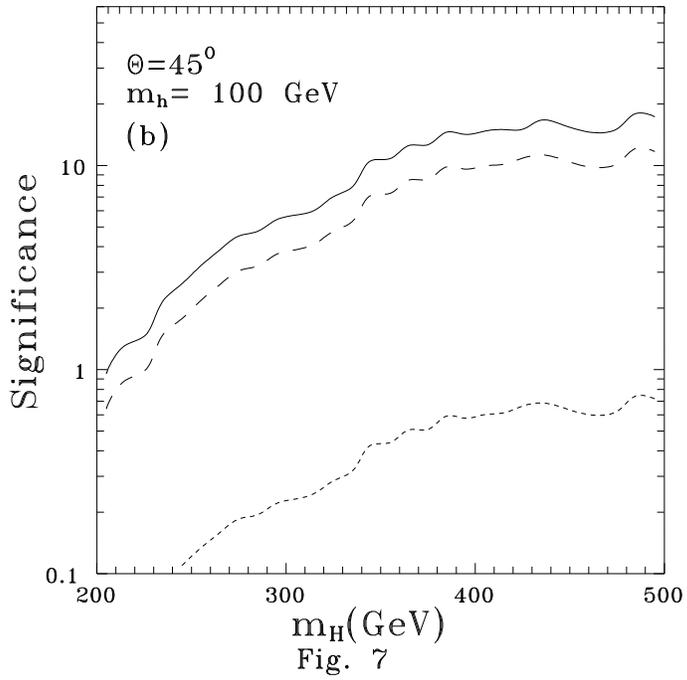
\end{document}